\documentstyle[12pt]{article}

\begin{document}
\newcommand {\be}{\begin{equation}}
\newcommand {\ee}{\end{equation}}
\newcommand {\bea}{\begin{array}}
\newcommand {\cl}{\centerline}
\newcommand {\eea}{\end{array}}
\renewcommand {\theequation}{\thesection.\arabic{equation}}
\newcommand {\newsection}{\setcounter{equation}{0}\section}
\def\nct{noncommutative torus }
\def\ncy{noncommutativity }
\def\nc{noncommutative }
\def\quan{quantization }
\def\BC{boundary condition }
\def\BCs{boundary conditions }
\def\con{constraints }
\def\Dcon{Dirac constraints }
\def\reg{regularization }
\def\dis{discretization }
\def\consys{constrained systems }
\def\e.o.m{equations of motion }
\def\Ham{Hamiltonian }
\def\Lag{Lagrangian }
\def\eps{\epsilon}
\baselineskip 0.65 cm
\begin{flushright}
IPM/P-99/037 \\
hep-th/9907055
\end{flushright}
\begin{center}
 {\Large {\bf  Boundary Conditions as Dirac Constraints}}
\vskip .5cm

M.M. Sheikh-Jabbari$^a$ and A. Shirzad$^{a,b}$
\footnote{ E-mails:jabbari@theory.ipm.ac.ir,  shirzad@cc.iut.ac.ir} \\

\vskip .5cm

 {\it $^a$ Institute for Studies in Theoretical Physics and Mathematics 
IPM,

 P.O.Box 19395-5531, Tehran, Iran}\\
{\it and}
\\
{\it $^b$ Department of Physics Isfahan University of Technology

Isfahan, Iran}
\end{center}

\vskip 2cm
\begin{abstract}
In this article we show that \BCs can be treated as \Lag and \Ham constraints.  
Using the Dirac method, we find that \BCs are equivalent to an infinite 
chain of second class \con which is a new feature in the context of constrained 
systems. Constructing the Dirac brackets and the reduced phase space structure 
for different boundary conditions, we show why mode expanding and then 
quantizing a field theory with \BCs is the proper way. We also show that in a 
quantized field theory subjected to the mixed boundary conditions, the field 
components are noncommutative. 

\end{abstract}

\vskip 2cm
PACS: 11.10.Ef, 11.25.-w, 04.60.Ds, 

Key words: Boundary conditions, Constraints, Dirac bracket.

\newpage
\section{Introduction}
\setcounter{equation}{0} 

It is well-known that to formulate a general classical field theory defined
in a box, besides the equations of motion one should  
know the behaviour of the fields on the boundaries, boundary conditions. 
Boundary conditions are usually relations between the fields and their various derivatives, 
including the time derivative, on the boundaries, expected to be held at all
the times.
In Hamiltonian language the \BCs are in general functions of the fields and their
conjugate momenta; hence the field theories subjected to the \BCs might be 
understood by the prescription for handling the \consys proposed by Dirac 
\cite{Dir}. 

In the usual field theory arguments, since \BCs are usually linear 
combinations of fields and their momenta, one can easily impose them on the 
solutions of the equations of motion, and find the final result. 
But, imposing the \BCs in some special cases 
may lead to inconsistencies with the canonical commutation relations
\cite{{AAS1},{AAS2},{CH1},{AAS3},{CH2},{Sei}}. 

In this article, considering the \BCs as constraints, we apply the Dirac's 
procedure to this constrained system. Although this idea have been used in 
\cite{{AAS3},{CH2}}, the problem has new and special features in the context of 
\consys on which, we mostly concentrate.

In the second section, we review the Lagrangian and Hamiltonian
constrained systems.
In section 3, to visualize the seat of \BCs we take a toy model and by 
discretizing the model show that \BCs are in fact the equations of motion for 
the points at the boundaries so that, when we go to the continuum limit, i.e. ,
the original theory, the acceleration term disappears. In other words \BCs 
are \Lag \con which are not consequences of a singular Lagrangian. 
In section 4, going to Hamiltonian picture we study the constraint structure
resulting from the boundary conditions, and apply it explicitly to some
field theories. 
Implying constraint consistency we show that although the Lagrange
multiplier is determined, the constraint chain is not terminated, 
This is a new feature in the \consys analysis. Exhausting all the consistency
checks we end up with an infinite constraint chain which all of them
are of second class, which is another new feature of this constraint structure. 
Moreover, we construct the fundamental Dirac brackets, the Dirac brackets of  
fields and their conjugate momenta.
In section 5, by a canonical transformation we go to the Fourier modes in 
terms of which, the constraint chain obtained in the previous section can be
easily solved. In this way we prove that, using the proper mode expansions 
is equivalent to working in the {\it reduced phase space}. 
In section 6, we apply the machinery developed in
the previous sections to the case of mixed boundary conditions, i.e. , we find 
the \con chain, the Dirac bracket and the reduced phase space.
The new and interesting result of this case is that, the Dirac bracket of two 
field components is obtained to be non-zero and hence, in the quantum theory 
these field components are noncommuting.
The last section is devoted to the concluding remarks.
\section{Review of Dirac Procedure}
\setcounter{equation}{0} 
Given the Lagrangian $L(q,\dot{q})$ (or $L(\phi,\partial\phi)$ in a field 
theory), the Lagrangian  equations of motion are: 
\be
{\bf L}_i=W_{ij}{\ddot{q}}_j +\alpha_i=0,
\ee
where ${\bf L}_i$ are Eulerian derivatives, $W_{ij}(q,\dot{q})\equiv{\partial^2
L \over \partial\dot{q}^i\partial\dot{q}^j}$ is called the Hessian matrix,
and $\alpha_i\equiv{\partial L \over \partial{q}^i}-\dot{q}_j({\partial^2 L 
\over \partial\dot{q}^i\partial{q}^j})$. If $|W_{ij}|=0$, the Lagrangian is 
called {\it singular} and in 
this case the number of equations containing accelerations are less than the 
number of degrees of freedom. Hence a number of Lagrangian constraints,
$\gamma^a(q,\dot{q})=0$, emerges (To obtain these  \con we should 
simply multiply both sides of (2.1) by the null eigenvector $\lambda^a_i$ of
$W$, so $\gamma^a(q,\dot{q})=\lambda^a_i\alpha_i$, \cite{Shir}.). Then we 
should add the time derivatives of
constraints,$\dot{\gamma}^a(q,\dot{q})$, to
the set of 
equations of motion to get new relations containing the accelerations. 
As a result two cases may happen
\newline
1) the rank of equations with respect to acceleration is equal to the number of 
degrees of freedom. 
\newline
2) new constraints, acceleration free relations, emerging.

In the first case the \e.o.m can be solved completely, however, the solutions 
should obey the acceleration free equations, the constraints. In the second,  
the derivatives of new \con and derivatives of previous 
\con should be added to the equations of motion, and the same scenario should be
repeated.

At the end, there may remain a number of undetermined accelerations;
it is shown that they correspond to the gauge degrees of freedom
and are related to the first class Hamiltonian constraints.
Moreover roughly speaking, there may exist some degrees of freedom which have
no dynamics and are completely determined via the constraints. These are related 
to the the second class Hamiltonian \con \cite{Gom}.

Let us study the Hamiltonian formulation.  Singularity of the Hessian matrix, 
${p_i \over \partial\dot{q}^j}$, implies the Legendre transformation, 
$(q,\dot{q})\rightarrow (q,p)$, to have a zero Jacobian and hence,
the set of  momenta, $p_i$; 
\be
p_i={\partial L \over \partial\dot{q}^i},
\ee
are not independent functions of $q$ and $\dot{q}$. So a number of 
Hamiltonian primary \con turns up
\be
\Phi^{(0)}_a(q,p)=0.
\ee

It can be shown that \cite{Dir} dynamics of any function in phase space is
obtained by
\be
{\dot g}\approx \{ g, H_T\}_{P.B.},
\ee
where weak equality, $\approx$, is the equality on the constraint surface, and
\be
H_T=H+\lambda_a\Phi_a,
\ee
is the total Hamiltonian, $\lambda_a$ being the Lagrange multipliers.

Like the \Lag case the consistency conditions of the primary \con should be 
investigated, i.e. the \con should be valid under the time evolution:
\be
{\dot \Phi^{(0)}_a}\approx \{ \Phi^{(0)}_a, H_T\}_{P.B.}\approx
\{\Phi^{(0)}_a,H \}+\lambda_b\{\Phi^{(0)}_a,\Phi^{(0)}_b\}\approx 0.
\ee
If the above relation dose not hold identically, then two possibilities 
remains
{\it i)} $\{\Phi^{(0)}_a,\Phi^{(0)}_b\}$'s vanish weakly. In this case new 
\Ham \con
\be
\Phi^{(1)}=\{\Phi^{(0)}_a,H \},
\ee
turns up.

{\it ii)} $\{\Phi^{(0)}_a,\Phi^{(0)}_b\}$ do not vanish, yielding equations 
for determining $\lambda_a$.

In general, depending on the rank of the matrix
$\{\Phi^{(0)}_a,\Phi^{(0)}_b\}$, 
we may have a mixture of two possibilities. That is, some of the Lagrange 
multipliers are determined and a number of new \con emerge. Here we do not 
bother the reader with the details. A complete and detailed discussion can be 
found in \cite{Gom}. 

Now the consistency conditions of $\Phi^{(1)}_a$ should be verified which
may result into some new \con $\Phi^{(2)}_a$. The procedure goes on, 
and finally we end up with some {\it constraint chains}. Roughly speaking, 
each chain
terminates if a Lagrange multiplier is determined or, if we get an
identically satisfied relation.
The latter case happens when the last constraint has weakly vanishing Poisson bracket
with the primary \con and the Hamiltonian.

We denote the set of \con $\Phi^{(1)},\Phi^{(2)},....$ as {\it secondary \con}.
These are really consequences of primary \con while the primary constraints, 
by themselves have their origin in the singularity of the \Lag 
(singularity of the Hessian matrix).
In a pure \Ham point of view, however, the origin   
of primary \con is not of essential importance. In any way given some
primary constraints, we should build the total Hamiltonian, (2.5), and
check their consistency.

There is another important classification of constraints: 
If the Poisson bracket of some constraint with all the \con in the chain 
vanishes, it is called {\it first class}. 
And if the matrix of mutual Poisson brackets of a subset of constraints, 
$C^{MN}$, 
\be          
C^{MN}=\{\Phi^{M},\Phi^{N}\},
\ee
has the maximally rank, it is invertible, we deal with 
{\it second class constraints}. It is shown that a constraint chain terminating 
with an identity, is of first class and ending with determining Lagrange 
multipliers are of second class \cite{Gom}.
To find the dynamics of a system with second class constraints, 
one may use the Dirac bracket,
\be
\{A,B\}_{D.B.}=\{A,B\}_{P.B.}-\{A,\Phi_M\}_{P.B.}\;(C^{-1})^{MN}
\{\Phi_N,B\}_{P.B.}.
\ee
The important property of the Dirac bracket is that for an arbitrary function 
$A$ and for all second class constraints $\Phi_M$,
\be
\{\Phi_M,A\}_{D.B.}=0.
\ee
It can be shown that, using the Dirac brackets instead of Poisson bracket, 
is equivalent to priory putting the second class \con {\it strongly} equal to 
zero.

For second class constraints we can always find a {\it canonical transformation} 
such that the constraints, $\Phi_M$, lie on the first $2n$ coordinates
$(q_1,...,q_n;p_1,...,p_n)$ of the phase space and the remaining 
degrees of freedom, $(Q_1,...,Q_{N-n};P_1,...,P_{N-n})$ are unconstrained.
The Dirac bracket in the original phase space is equal to the Poisson bracket 
in the space $(Q_1,...,Q_{N-n};P_1,...,P_{N-n})$, the {\it reduced phase space}
\cite{{Dir},{Wein},{Mas}}.
Although finding the above canonical transformation is not an easy task,
for the case we study in this paper, \BCs as constraints, we show that using the
suitable mode expansions, is in fact equivalent to going to reduced phase space.

\section{Boundary Conditions as Constraints}
\setcounter{equation}{0} 
Boundary conditions are acceleration free equations which in general are not
related to a singular Lagrangian. To visualize this point, let us take a simple
(1+1) field theory as a toy model
\be
S={1\over2}\int_0^l dx\int_{t_1}^{t_2}dt \;\;\bigl[(\partial_t\phi)^2-
(\partial_x\phi)^2\bigr].
\ee
Variation of the action with respect to $\phi$ gives
\be
\delta S=\int_0^l dx\int_{t_1}^{t_2}dt {\bf L}(\phi) \delta\phi + 
\int_{t_1}^{t_2}dt (\partial_x\phi)\delta\phi|_0^l+
\int_{0}^{l}dt (\partial_t\phi)\delta\phi|_{t_1}^{t_2},
\ee
where ${\bf L}(\phi)=\partial^2_t\phi-\partial^2_x\phi$, is the Eulerian 
derivative. For an arbitrary $\delta\phi$, variation of the action vanishes if the three 
terms in the above equation vanish independently. 
The first term in (3.2) leads to \e.o.m and the last term to the initial 
conditions. The second term which is called the surface term, results in the
boundary conditions. For this term to vanish, there are two choices
$\delta\phi|_{boundary}=0$, Dirichlet boundary conditions, or 
$\partial_x\phi|_{boundary}=0$, Neumann boundary conditions. 
The \BCs unlike the equations of motion, are acceleration-free 
equations and should be held at all the times. In other words, they can be 
treated as \Lag constraints. To clarify
this point we repeat the above argument in the discrete version:
\be
S={1\over2}\int_{t_1}^{t_2}dt \sum_{i=0}^{N}
\eps(\partial_t\phi_i)^2-
\sum_{i=0}^{N-1}{1\over \eps}(\phi_i-\phi_{i+1})^2,
\ee
\be
\phi_i(t)=\phi(x,t)|_{x=x_i}\;\;\;\ ; x_n=n\eps,
\ee
and $\eps={l\over N}$ so that $\eps\rightarrow 0$ ($N\rightarrow \infty$) 
reproduces the continuum theory.

Demanding the variation of (3.3) to vanish, leads to 
\footnote{ It is worth noting that we still have the option $\delta\phi_0$ or
$\delta\phi_N=0$ which in the continuum limit translate into the Dirichlet \BCs.} 
\be
\eps\partial^2_t\phi_0={1\over \eps}(\phi_1-\phi_0),
\ee
\be
\eps\partial^2_t\phi_i={1\over \eps}(\phi_{i+1}-2\phi_i+\phi_{i-1}),\;\; 
i\neq0,N
\ee
\be
\eps\partial^2_t\phi_N={1\over \eps}(\phi_N-\phi_{N-1}).
\ee
Taking the continuum limit, assuming that acceleration of the end point are 
finite,  the equations for $0,N$ give
\be
\lim{1\over \eps}(\phi_1-\phi_0)=0 \;\;\ {\rm and} 
\;\;\;\; \lim{1\over \eps}(\phi_N-\phi_{N-1})=0. 
\ee
Hence in the continuum limit {\it \e.o.m} for the end points give acceleration
free equations, the \Lag constraints, where as (3.6) leads to ${\bf L}(\phi)=0$,
which actually contains the acceleration term.

A new feature appearing here is that, unlike
the usual \Lag constraints, \BCs are the \con which are not consequences of 
the singularity of Lagrangian, but a result of taking the continuum limit.
\section{The \Ham Setup}
\setcounter{equation}{0} 
In this section, by going to \Ham formulation, we apply the Dirac procedure
to a field theory with given boundary conditions. Again, we take our simple toy model and 
treat the \BCs as \Ham primary constraints:
\be
\Phi^{(0)}=\partial_x\phi|_{x=0}.
\ee
Here we explicitly work out Neumann \BC at one end , 
the Neumann \BC at the other end and the Dirichlet cases can be worked out 
similarly. 
The total \Ham is built by adding the constraint to the \Ham by arbitrary 
Lagrange multiplier
\be
H_T=H+\lambda\Phi^{(0)},
\ee
with
\be
H={1\over 2}\int_0^l dx \;\;\;\Pi^2+(\partial_x\phi)^2, 
\ee
\be
\Pi=\partial_t\phi.
\ee
We should remind that as discussed in sec. 2, the appearance of the \con
(4.1) is not a consequence of the definition of the momenta for an ordinary 
singular Lagrangian and hence, the transformation (4.4) between the velocities 
and momenta is well-defined and invertible throughout all the points, even at 
the boundaries.

Now we should check the consistency condition
\be
{\dot \Phi^{(0)}}=\{ \Phi^{(0)}, H_T\}_{P.B.}=\partial_x\Pi|_0\equiv\Phi^{(1)},
\ee
which leads to the secondary constraint, $\Phi^{(1)}$. It should be noted that
to obtain (4.5), although the conditions are imposed at the boundaries, 
the fields can safely be extended into neighbourhood of the boundaries and we can
use $\Phi^{(0)}=\int \delta (x) \partial_x\phi dx$.

We should go further:
\be
{\dot \Phi^{(1)}}=\{\Phi^{(1)}, H_T\}=
\{\Phi^{(1)},H \}+\lambda\{\Phi^{(1)},\Phi^{(0)}\}=0.
\ee
The second term on the right hand side, 
\be\bea{cc}
\lambda\{\Phi^{(1)},\Phi^{(0)}\}=
\int\delta(x)\delta(x')\{\partial_x\Pi, \partial_{x'}\phi\} dxdx'\\ \;\;\;\;\; 
\;\;\;\;\;=-\int\delta(x)\delta(x')\partial_x\partial_{x'}\delta(x-x') dxdx',
\eea\ee
is not well-defined, and formally can be written as 
$\partial^2_{x}\delta(x-x')|_{x=x'=0}$. 
This term compared to the first term is infinitely large. The only way to impose
the consistency condition on the \con is
\be
\lambda=0,
\ee
and
\be
\{\Phi^{(1)},H \}=0.
\ee
There is a new feature appearing which is not any of the cases {\it i)} and 
{\it ii)} discussed  in section 2. The consistency condition,
(4.6), reduces to two equations, (4.8) and (4.9), and although {\it the 
Lagrange multiplier is determined the constraint chain is not terminated}.

The above discussion can be better understood if the calculation is regularized
by considering the discrete case. Using the the discrete version of equation
(4.6), $\lambda$ is turned out to be of the order of $\eps$, going
to the continuum limit is vanishes, and the other term, 
$\{\Phi^{(1)},H \}$, should vanish separately.

Defining $\{\Phi^{(1)},H \}$ as $\Phi^{(2)}$, the other secondary constraint, 
we find
\be
\Phi^{(2)}=\partial_x^3\phi|_0.
\ee
We should go on:
\be
\Phi^{(3)}\equiv{\dot \Phi^{(2)}}=\{\Phi^{(2)}, H_T\}=\{\Phi^{(2)},H \}=
\partial_x^3\Pi|_0.
\ee
This process should be continued and finally we are left with an infinite 
number of constraints:
\be
\Phi^{(n)}=\left\{\bea{cc}
\partial_x^{(n+1)}\phi|_0 \;\;\;\; n=0,2,4,... \\
\partial_x^{(n)}\Pi|_0 \;\;\;\; n=1,3,5,... 
\eea\right.
\ee
Exhausted the constraint consistency conditions, 
we show that the Poisson bracket of the constraints,
\be
C_{mn}\equiv\{ \Phi^{(m)}, \Phi^{(n)}\},
\ee
is non-singular and hence, the set of \con (4.12) are all of second class. To 
show this first we calculate
\be
C_{mn}=\left\{\bea{cc}
0 \hspace{3cm} m,n=0,2,4,... \\
0 \hspace{3cm} m,n=1,3,5,... \\
\int\delta(x)\delta(x')\partial^{m+1}_x\partial^n_{x'}\delta(x-x')dxdx'
\;\;\;\; m=0,2,4,...., n=1,3,5,... 
\eea\right.
\ee
To find $det C$, the non-zero elements should be regularized. This \reg 
can be done by two methods, \dis or using a limit of a regular function, e.g. 
the Gaussian function, to represent $\delta(x)$. Here we choose the second, but
one can easily show that the other method gives the same results. Inserting
\be
\delta(x-x')=\lim_{\eps\rightarrow 0}\;\; {1\over \eps\sqrt\pi} e^{{-(x-x')^2 
\over \eps^2}}
\ee
into (4.14) we find
\be\bea{cc}
\int\delta(x)\delta(x')\partial^{m+1}_x\partial^n_{x'}\delta(x-x')dxdx'=
{-1\over \sqrt\pi} \eps^{-(m+n+2)} H_{m+n+1}(0) \\
\;\;\;\;\;\;\;\;\;\
={-1\over \sqrt\pi}(-2)^{(n+m+1)/2} \eps^{-(m+n+2)} (m+n)!!,
\;\;\;\; m=0,2,4,..., n=1,3,5,... 
\eea\ee
$H_n(0)$ denotes the Hermite polynomials at $x=0$ \cite{Hand}.
Putting these together, $C$ is finally found to be
\be
C=A\otimes B,
\ee
where
$$
A=\left(\bea{cc} 0\;\;\;\ 1\\
-1\;\;\ 0
\eea\right),
$$
and $B$ is an infinite dimensional matrix with
\be
B_{mn}={-1\over \sqrt\pi}(-2)^{(n+m-1)} {((2(m+n)-3)!!\over\eps^{-2(m+n)-1}}
\;\;\;\;\;\; m,n=1,2,....
\ee
It is straightforward to show that the matrix $B$ has non-zero 
determinant, i.e. the matrix $C$ is invertible, and hence all the \con in the 
chain are second class. One way to consider all of them is using the Dirac 
bracket. To find Dirac bracket of any two arbitrary functions in the phase 
space, it is enough to calculate Dirac brackets of $(\phi,\phi)\;,\;(\phi,\Pi)\;{\rm and}
\; (\Pi,\Pi)$ 
\footnote{Find more detailed calculations in the appendix.}

\be
\{\phi(x),\phi(x')\}_{D.B.}=-\{\phi(x),\Phi^{(m)}\}C^{-1}_{mn} 
\{\Phi^{(n)},\phi(x')\}=0.
\ee
\be
\{\Pi(x),\Pi(x')\}_{D.B.}=-\{\Pi(x),\Phi^{(m)}\}C^{-1}_{mn} 
\{\Phi^{(n)},\Pi(x')\}=0.
\ee
\be\bea{cc}
\{\phi(x),\Pi(x')\}_{D.B.}=\delta(x-x')-\{\phi(x),\Phi^{(m)}\}C^{-1}_{mn} 
\{\Phi^{(n)},\Pi(x')\} \\ \;\;\;\;\; 
=\delta(x-x')-R(x,x').
\eea\ee
Without using the explicit form of $C^{-1}$ one can show
\be
R(x,x')=\kappa\eps\delta(x)\delta(x'),
\ee
where $\kappa$ is a numeric factor. To find $\kappa$, let us
obtain the Dirac bracket of the constrain $\Phi^0$ with an arbitrary function 
$f$, using (2.10) we should have
\be
\{\partial_x\phi(x)|_0,f(\phi,\Pi)\}_{D.B.}=
\int\delta(x)\partial_x\{\phi(x),f\}_{D.B.}=0.
\ee
Denoting ${\partial f\over \partial (\Pi(x'))}\equiv g(x')$, we can write
\be
\int\delta(x)\partial_x\{\phi(x),\Pi(x')\}_{D.B.}g(x')=0.
\ee
Inserting (4.21) and (4.22) into (4.24) reduces to
\be
\int(\partial_x\delta(x)+\kappa\eps\delta(x)\partial_x\delta(x))g(x)=0.
\ee
Remembering (4.15), we find
\be
\kappa=-\sqrt\pi.
\ee
Hence
\be
\{\phi(x),\Pi(x')\}_{D.B.}=\delta(x-x')+\kappa\eps\delta(x)\delta(x').
\ee
Appearance of \reg parameter, $\eps$, in the Dirac bracket sounds bad, but 
since the second term has two delta functions, to be of the same order
of the first term, in fact an $\eps$ factor is necessary. We will clarify and 
discuss this point in the next section.

The Dirichlet \BC can be worked out similarly. In this case the constraint 
chain is obtained to be
\be
\Phi^{(n)}=\left\{\bea{cc}
\partial_x^{(n)}\phi|_0 \;\;\;\; n=0,2,4,... \\
\partial_x^{(n-1)}\Pi|_0 \;\;\;\; n=1,3,5,... 
\eea\right.
\ee
Performing the calculations, one can show that the Dirac brackets are like
the Neumann case, except for 
the $\kappa$ factor, which is $+\sqrt\pi$.
\section{Mode expansion and Reduced Phase Space}
\setcounter{equation}{0} 
In the previous section we showed that a field theory subjected to the Neumann 
or Dirichlet \BCs is a system constrained to an infinite chain of {\it second
class} constraints. As mentioned in sec. 2, for a system with second class 
constraints, there is a subspace of phase space which is spanned by a set of 
unconstrained canonical variables, the reduced phase space. The important 
property of these variables is that,
Poisson bracket in terms of them is equivalent to the Dirac bracket defined 
on the whole constrained phase space. 

In this section we will explicitly find the reduced phase space and show that
it is in fact equivalent to phase space determined by the Fourier modes.

Let us consider the Fourier transformed variables
\be\bea{cc}
\phi(x)={1\over\sqrt{2\pi}}\int \phi(k)e^{ikx}dk\;\;\;\; , \;\;\;\;\;
\phi(k)={1\over\sqrt{2\pi}}\int \phi(x)e^{-ikx}dx \\
\Pi(x)={1\over\sqrt{2\pi}}\int \Pi(k)e^{-ikx}dk\;\;\;\; , \;\;\;\;\;
\Pi(k)={1\over\sqrt{2\pi}}\int \Pi(x)e^{ikx}dx. 
\eea\ee
One can easily show that the above transformation is canonical:
\be
\{\phi(k),\phi(k')\}=0\;\; ,\;\;\ 
\{\Pi(k),\Pi(k')\}=0\;\;\; , \;\;\ 
\{\phi(k),\Pi(k')\}=\delta(k-k').
\ee

The Neumann (Dirichlet) constraint chain, (4.12) and (4.28), in terms of the new variables are easily  
obtained:
All the odd (even) moments of $\phi(k)$ and $\Pi(k)$ are zero.
The most general solution to these conditions is that $\phi(k)$ and $\Pi(k)$ are even
(odd) functions of $k$. Then (5.1) gives
\footnote{For Dirichlet case Cosine should be replaced be Sine.}
\be
\phi(x)={1\over\sqrt{\pi}}\int \phi(k)\cos{kx}dk\;\;\;\; , \;\;\;\;\;
\Pi(x)={1\over\sqrt{\pi}}\int \Pi(k)\cos{kx}dk.
\ee
The main advantage of the Fourier modes, $\phi(k)$ and $\Pi(k)$, 
is that although they are limited to even (odd) functions, are still
canonical variables; in contrast with the original fields  
$\phi(x)$ and $\Pi(x)$ which lose their usual canonical structure due to 
constraints.

To compare the Dirac bracket results with those of reduced phase space, we
work out Poisson brackets of $\phi(x)$ and $\Pi(x)$. Using (5.2) and (5.3) 
we have
\be\bea{cc}
\{\phi(x),\phi(x')\}=0,\;\; \\
\{\Pi(x),\Pi(x')\}=0, \\
\{\phi(x),\Pi(x')\}={1\over{\pi}}\int\cos{kx}\cos{kx'}dk\equiv\delta_N(x,x'),
\eea\ee
for Neumann boundary conditions. For Dirichlet case only $\{\phi,\Pi\}$ differs from 
above:
\be
\{\phi(x),\Pi(x')\}={1\over{\pi}}\int\sin{kx}\sin{kx'}dk
\equiv\delta_D(x,x').
\ee
Performing the integrations we have
\be\bea{cc}
\delta_N(x,x')=\delta(x-x')+\delta(x+x'), \\
\delta_D(x,x')=\delta(x-x')-\delta(x+x').
\eea\ee
If we consider only the positive $x$'s, $x\geq 0$, $\delta_N$ and $\delta_D$
for $x,x'\neq 0$ are exactly $\delta(x-x')$. For $x,x'=0$, we regularize 
delta functions by
\be
\left\{\bea{cc}
\delta(x-x')+\delta(x+x')={2\over\sqrt{\pi}\eps}, \\
\delta(x-x')-\delta(x+x')=0.
\eea\right.
\hspace {2cm} {\rm at} \;\; x=x'=0
\ee
Hence $\delta_N$ and $\delta_D$ for $x\geq 0$ is in exact agreement with the 
Dirac bracket results obtained in previous section.
The above argument clarifies, why using the usual mode expansions to quantize a
system with Neumann or Dirichlet boundary condition, i.e. , imposing the \BCs 
and {\it then} quantizing, works.
\section{Mixed Boundary Conditions, Another Example}
\setcounter{equation}{0} 
In this section we handle a more general family of boundary conditions, 
mixed boundary conditions, which are combinations of Neumann and Dirichlet 
cases. It has been shown that  
these \BCs lead to unusual results in the context of string theory
\cite{{AAS1},{AAS2},{CH1},{AAS3},{CH2},{Sei}}.

As a toy model for a field theory resulting in the mixed \BCs consider
\be
S={1\over2}\int_0^l dx\int_{t_1}^{t_2}dt\big[(\partial_t\phi_i)^2-
(\partial_x\phi_i)^2+ F_{ij}\partial_t\phi_i\partial_x\phi_j\bigr],
\ee
where $i,j=1,2$ and $F_{ij}$ is a constant antisymmetric background.
Varying $S$ with respect to $\phi_i$, gives:
\be
\;\;\; \partial^2_{t}\phi_{i}-\partial^2_{x}\phi^{i}=0,
\ee
\be
\partial_{x}\phi_{i}+{\cal F}_{ij}\partial_{t}\phi_{j}=0 \;\;\;
{\rm at}\;\; x=0,l.
\ee
Equations (6.3), as discussed in section 3, give the \Lag constraints.
In the discretized version, (6.3) are the \e.o.m for the end points and
in the continuum limit, the acceleration term disappears. 
It is worth noting that (6.3) reproduce the Neumann and Dirichlet \BCs 
for $F=0$ and $\infty$ respectively.

Now to apply the Dirac method, we go to \Ham formulation:
\be
\Pi_i=\partial_{t}\phi_{i}+{\cal F}_{ij}\partial_{x}\phi_{j},
\ee
\be
H= {1 \over 2} \int_0^l(\Pi_i-F_{ij}\partial_{x}\phi_j)^2
+(\partial_{x}\phi_i)^{2}\;\; dx,
\ee
and the primary constraints,
\be
\Phi_i^{(0)}=\Phi_i(x)|_{x=0},
\ee
with
\be
\Phi_i(x)\equiv M_{ij}\partial_{x}\phi_{j}+{\cal F}_{ij}\Pi_{j}=0\;\; , 
\;\; {M}_{ij}=({\bf 1}- {F}^2)_{ij}.
\ee
Note that in this case the \Lag constraints, (6.3), depends on velocities, 
and as mentioned before, the transformation (6.4), is non-singular and the 
\Lag \con can be translated into \Ham constraints, (6.7), without any 
difficulty. The consistency of the primary \con should be verified:
\be
{\dot \Phi_i^{(0)}}=\{\Phi_i^{(0)}, H_T\}=
\{\Phi_i^{(0)},H \}+\lambda_j\{\Phi_i^{(0)},\Phi_j^{(0)}\}=0.
\ee
The first term is easy to work out: 
\be
\Phi_i^{(1)}=\{\Phi_i^{(0)},H \}=\partial_x\Pi_i|_{x=0}.
\ee
Similar to the arguments of sec. 4, $\{\Phi_i^{(0)},\Phi_j^{(0)}\}$ is 
infinitely large compared to the first term, and the only way for (6.8) to be satisfied is
\be
\lambda_i=0\;\;\;\;\;{\rm and} \;\;\;\; \Phi_i^{(1)}=0.
\ee
Again, although the Lagrange multiplier, $\lambda_i$,
is determined, there are secondary constraints, $\Phi_i^{(1)}=0$. Moreover we have
the advantage that $\lambda_i$ disappears in the remaining steps.

Direct calculations on the consistency conditions for the \con leads to the 
chain
\be
\Phi_i^{(n)}=\left\{\bea{cc}
\partial_x^{n}\Phi_i|_0 \;\;\;\; n=0,2,4,... \\
\partial_x^{(n)}\Pi_i|_0 \;\;\;\; n=1,3,5,... 
\eea\right.
\ee
To verify that these \con are really second class, we study the matrix,
$C_{ij}^{mn}\equiv\{ \Phi_i^{(m)}, \Phi_j^{(n)}\}$:
\be
C_{ij}^{mn}=\left\{\bea{cc}
0 \hspace{4cm} m,n=1,3,5,... \\
-2(MF)_{ij}\int \delta(x) \delta(x')
\partial^{m+1}_x\partial^n_{x'}
\delta(x-x') dxdx' 
\;\;\;\; m,n=0,2,4,... \\
M_{ij}\int\delta(x)\delta(x')
\partial^{m+1}_x\partial^n_{x'}
\delta(x-x')dxdx'
\;\;\;\; m=0,2,4,...., n=1,3,5,... 
\eea\right.
\ee
$C$ can be written in the form of
\be
C=F\otimes B,
\ee
where $F$ is a $4\times4$ matrix:
\be
F=\left(\bea{cc} -2(MF)\;\;\;\ M\\
-M\hspace{1.0cm} 0
\eea\right),
\ee
and $B$ given by (4.18). In section 4, we discussed that $B$ is invertible.
Since $det\;F\neq0$, $C$ is invertible too, hence all the \con in the  
chain (6.11) are second class. 

One can show that the fundamental Dirac brackets are as following
\be\bea{cc}
\{\phi_i(x),\phi_j(x')\}_{D.B.}=-\{\phi_i(x),\Phi_k^{(m)}\}(C^{-1})_{kl}^{mn} 
\{\Phi_l^{(n)},\phi_j(x')\}\\ \hspace{2.1cm}
=(-2M^{-1}F)_{ij}(\eps^2\sqrt{\pi}\delta(x)\delta(x'))
\eea\ee
\be
\{\Pi_i(x),\Pi_j(x')\}_{D.B.}=-\{\Pi_i(x),\Phi_k^{(m)}\}(C^{-1})_{kl}^{mn}
\{\Phi_l^{(n)},\Pi_j(x')\}=0.
\ee
\be\bea{cc}
\{\phi_i(x),\Pi_j(x')\}_{D.B.}=\delta(x-x')-\{\phi_i(x),\Phi_k^{(m)}\}
(C^{-1})_{kl}^{mn}\{\Phi_l^{(n)},\Pi_j(x')\} \\ \;\;\;\;\; 
\;=\delta(x-x')-R(x,x')=\delta_N(x,x').
\eea\ee

The important result of the mixed case is (6.15); the Dirac bracket of two
field
components are non-zero. This means that in the quantized theory these field
components are noncommuting. In the string theory, where the fields describe
the space coordinates, (6.15) tells us that, the space probed by open strings 
with mixed boundary conditions is a {\it noncommutative} space \cite{{AAS1},
{AAS2},{CH1}}.

Using the canonical (or Fourier) transformations, (5.1) and (5.2), we can explicitly
build up the reduced phase space for the mixed case.
Let $\Phi_i(k)$ represent the Fourier modes of $\Phi_i(x)$ defined in (6.7),
\be
\Phi_i(x)={1\over\sqrt{2\pi}}\int \Phi_i(k)e^{ikx}dk\;\;\;\; , \;\;\;\;\;
\Phi_i(k)={1\over\sqrt{2\pi}}\int \Phi_i(x)e^{-ikx}dx,
\ee
Using (5.2), Poisson brackets of $\Phi_i(k)$ and $\Pi_i(k)$ can be worked out.
Imposing the \con (6.11), we find that $\Phi_i(k)$ and $\Pi_j(k)$, are odd
and even functions of $k$ respectively:
\be
\Phi_i(x)={1\over\sqrt{\pi}}\int \Phi_i(k)\sin{kx}dk\;\;\;\; , \;\;\;\;\;
\Pi_i(x)={1\over\sqrt{\pi}}\int \Pi_i(k)\cos{kx}dk.
\ee

Remembering (6.7), we can derive the field components:
\be
\phi_i(x)={M^{-1}_{ij}\over\sqrt{\pi}}\int {-dk\over k}(\Phi_j(k)
\cos{kx}+F_{jk} \Pi_k(k)\sin{kx}),
\ee
which explicitly satisfies the mixed boundary conditions.

Derived the mode expansions of the fields and their conjugate momenta, 
we finds their Poisson brackets:
\be\bea{cc}
\{\phi_i(x),\phi_j(x')\}={1\over{\pi}}\int {dk\over k}{dk'\over k'}
\bigl[(M^{-1}F)_{ik}\{\Phi_k(k),\Pi_l(k')\}M^{-1}_{lj}
\cos{kx}\sin{k'x'}+\\ \hspace{4.3cm}
+(M^{-1}F)_{jk}\{\Pi_k(k),\Phi_l(k')\}M^{-1}_{il}
\cos{k'x'}\sin{kx}\bigr] \\ \hspace{1.0cm}
={-1\over{\pi}}\int{dk\over k}(M^{-1}F)_{ij}(\cos{kx'}\sin{kx}+\cos{kx}\sin{kx'})
\\  =(M^{-1}F)_{ij}\int^x(\delta_N(y,x')-\delta_D(y,x'))dy\\
=-2(M^{-1}F)_{ij}\int^x\delta(y+x')dy.
\eea\ee
Since for $x,x'\geq 0$
\be
\int^x\delta(y+x')dy=\left\{\bea{cc}1 \;\;\;\;\;\; x=x'=0 \\
0 \;\;\;\;\;\; otherwise,
\eea\right.
\ee
(6.21) is non-zero only for $x,x'=0$:
\be
\{\phi_i(0),\phi_j(0)\}=-2(M^{-1}F)_{ij}.
\ee
Comparing (6.21) and (6.15), we find that they are exactly the same.
In other words, (6.19) and (6.20) are functions defining the reduced phase 
space.

In the context of string theory, (6.21) implies that the end points of open 
strings subjected to mixed \BCs are living in a {\it noncommutative} space.
The mixed open strings appear when we are studying D-branes in a NSNS two-form
background. In this case, (6.21) tells us that the world-volume of such branes 
are noncommutative planes.

We can also calculate $\{\Pi_i(x),\Pi_j(x')\}$ and $\{\phi_i(x),\Pi_j(x')\}$.
The results are in exact agreement with (6.16) and (6.17).

\section{Concluding Remarks}
\setcounter{equation}{0} 
In this paper, we studied the old and well-known problem of field theories
with boundary conditions from a new point of view. We discussed that in the 
\Lag formulation \BCs are \Lag \con which are not a consequence of a singular 
Lagrangian. For further 
study we built the \Ham formulation, and considered \BCs as primary constraints. 
Asking for the \con consistency conditions we found two new 
features in the context of constrained systems

1) Although the Lagrange multiplier in the total \Ham is determined, the  
\con chain is continued.

2) Boundary conditions are equivalent to an {\it infinite} chain of 
{\it second class} constraints.

Constructing the Dirac brackets of the fields and their conjugate momenta for these
second class constraints, we showed that the method based on mode expansion, 
 is equivalent to working in the reduced phase space.

The relation between \Ham method we built here and the usual method of
imposing boundary conditions in the equations of motion, can simply be 
understood.
In the former, to ensure that \BCs are satisfied, we make the Taylor expansion 
of \BCs as a function of time, and put all the coefficient equal to zero.
These
coefficients are exactly our constraint chain. But, in the latter, the Fourier 
mode expansion is used and \BCs are guaranteed by choosing all the Fourier 
components to satisfy boundary conditions.

In the last section of the paper, we handled the mixed \BCs which seems to be 
an exciting problem in the context of string theory \cite{Sei}. Having 
noncommuting field components, is the interesting feature appearing in this case. 
Besides the string theory, mixed \BCs can be encountered in the context of electrodynamics
having an extra $\theta$-term:
$$
S={1\over4}\int ({\cal F}_{\mu\nu}^2+\theta\eps_{\mu\nu\alpha\beta}
{\cal F}_{\mu\nu}{\cal F}_{\alpha\beta}).
$$
In the above action $\theta$ plays a role similar $F$ in our toy model.
Varying the action gives a surface term, vanishing of which leads 
the mixed boundary conditions. Quantizing this theory is an interesting problem 
we postpone it to future works.

\vskip 1cm
{\bf \large Acknowledgements}
\newline
M.M Sh-J. would like to thank F. Ardalan and H. Arfaei for helpful discussions 
and also P-M. Ho for reading the manuscript.
\vskip 1cm
{\bf \large Appendix}: 
In this appendix we present some of the calculation details

$$
K^{(m)}(x)\equiv\{\phi(x),\Phi^{(m)}\}=\left\{\bea{cc}
0 \hspace{3cm} m=0,2,4,... \\
\{\phi(x),\partial_x^m\Pi^{(m)|_0}\}=k^m(x) \;\;\;\; m=1,3,5,... .
\eea\right.
$$
$$
L^{(m)}(x)\equiv\{\Pi(x),\Phi^{(m)}\}=\left\{\bea{cc}
\{\Pi(x),\partial_x^m\phi^{(m)|_0}\}=l^m(x) \;\;\;\; m=0,2,4,... \\
0 \hspace{3cm} m=1,3,5,... .
\eea\right.
$$
$$
k^m(x)=\int\partial^{m}_{x'}\delta(x-x')\delta(x')dx'=
{1\over \sqrt\eps\pi} exp{({-x^2\over\eps^2})}\;\;{1\over\eps^{m}} H_{m}(0)
\equiv\delta(x)k_m.
$$
$$
l^m(x)=-\int\partial^{m+1}_{x'}\delta(x'-x)\delta(x')dx'=
{1\over \sqrt\eps\pi} exp{({-x^2\over\eps^2})}\;\;{1\over\eps^{m+1}} 
H_{m+1}(0)\equiv\delta(x)k_{m+1}.
$$
where, $H_{m}(0)$ is the Hermite polynomial at zero.
Then one can easily work out 
\newline $\{\phi(x),\Pi(x')\}_{D.B.}$
$$
\{\phi(x),\Pi(x')\}_{D.B.}=\delta(x-x')+
k_{m+1}k_{n}B^{-1}_{mn}\delta(x)\delta(x')
$$ 
The power of $\eps$ in $k_{m+1}k_{n}B^{-1}_{mn}$, 
can be read off from the explicit form of $k_m$ and $B_{mn}$, and the results is 
$k_{m+1}k_{n}B^{-1}_{mn}=\kappa\eps$. 

Calculations for the mixed \BCs can be performed similarly.



\begin{thebibliography}{99}

\bibitem{Dir}
P.A.M. Dirac, "Lecture Notes on Quantum Mechanics", Yeshiva University New York, 
1964. Also see, P.A.M. Dirac, Proc. Roy. Soc. London, ser. A, {\bf 246}, 326 
(1950). 

\bibitem{AAS1} F. Ardalan, H. Arfaei, M. M. Sheikh-Jabbari,
"Mixed Branes and Matrix Theory on Noncommutative Torus", Proceeding of 
PASCOS 98, hep-th/9803067.

F. Ardalan, "String Theory, Matrix Model, and Noncommutative Geometry",  
hep-th/9903117.

\bibitem{AAS2} F. Ardalan, H. Arfaei, M. M. Sheikh-Jabbari,
"Noncommutative Geometry From Strings and Branes", JHEP 02 (1999) 016. 

\bibitem{CH1}
C.-S. Chu, P.-M. Ho, "Noncommutative Open Strings and D-branes", 
{\it Nucl. Phys.} {\bf B550} (1999) 151, hep-th/9812219.  

\bibitem{AAS3} F. Ardalan, H. Arfaei, M. M. Sheikh-Jabbari,
"Dirac Quantization of Open Strings and Noncommutativity in Branes", 
hep-th/9906161.

\bibitem{CH2}
C.-S. Chu, P.-M. Ho, "Constrained Quantization of Open Strings in
Background B Field and Noncommutative D-branes", hep-th/9906192.  

\bibitem{Sei}
N. Seiberg, talk given in the conference {\it New Ideals in Particle Physics and
Cosmology}, Uni. Penn. , May 19-22, 1999.

\bibitem{Shir} 
A. Shirzad, {\it Jour.} of {\it phys.}{\bf A}: MAth. Gen. vol.31 (1998),
2747.  

\bibitem{Gom}
C. Battle, J.M. Gomis, and N. Roman-Roy, {\it Jour.} {\it Math.} {\it
Phys.} vol. 27 (1986) 2953.

\bibitem{Wein}
S. Weinberg, "The Quantum Theory of Fields", vol. 1, Cambridge University Press.  

\bibitem{Mas}
T. Maskawa and H. Nakajima, "Singular Lagrangian and the Dirac-Faddeev Method",
Prog. Theo. Phys, Vol. 56, (1976) 1295. 

\bibitem{Hand}
Murray R. Spiegel, Mathematical Handbook, Schaum's outline series.

\end{thebibliography}
\end{document}